\documentclass[aps,prl,twocolumn,groupedaddress]{revtex4}

\usepackage{graphicx}

\usepackage{amsmath,amsthm,amsfonts, latexsym}

\newcommand{\mtx}[2]{\left(\begin{array}{#1}#2\end{array}\right)}

\begin{document}


\title{Entanglement Cost of Nonlocal Measurements}


\author{Somshubhro Bandyopadhyay$^1$, Gilles Brassard$^1$, Shelby Kimmel$^2$, and William K. Wootters$^2$}
\affiliation{$^1$D\'epartement IRO, Universit\'e de Montr\'eal, Montr\'eal, Qu\'ebec, H3C 3J7, Canada\\
$^2$Department of Physics, Williams College, Williamstown, MA 01267, USA}


\date{\today}

\begin{abstract}
For certain joint measurements on a pair of spatially separated particles, we ask how much
entanglement is needed to carry out the
measurement exactly.  For a class of orthogonal measurements on two qubits with 
partially entangled eigenstates, we present upper and lower bounds on the 
entanglement cost.  The upper bound is based on a recent result by D.~Berry [Phys.\ Rev.\ A {\bf 75}, 032349 (2007)].  The lower bound, based on the entanglement production capacity of the measurement,
implies that for almost all measurements in the class we consider, the entanglement required to perform the measurement is strictly greater than the average
entanglement of its eigenstates.  On the other hand, we show that for any complete measurement in 
$d \times d$ dimensions that is invariant under all local Pauli operations, the cost of the measurement is exactly equal to
the average entanglement of the states associated with the outcomes.  
\end{abstract}


\maketitle

\section{I. Introduction}

In this paper we ask how much entanglement is required to perform a measurement
on a pair of spatially separated systems, if the participants are allowed only local operations
and classical communication.  That is, we want to find the ``entanglement cost'' of a
given measurement.  (We give a precise definition of this term in the following subsection.)
Our motivation can be traced back to a 1999 paper
entitled ``Quantum nonlocality without entanglement'', which presents a complete
orthogonal measurement that cannot be performed using only local operations
and classical communication (LOCC), even though the eigenstates of the measurement
are all unentangled \cite{nwoe}.  That result shows that there can be a kind of nonlocality in a quantum 
measurement that is not captured by the entanglement of the associated states.  
Here we wish to quantify this nonlocality for specific measurements.  Though the measurements
we consider here have outcomes associated with {\em entangled} states, we find that the 
entanglement cost of the measurement often exceeds the entanglement
of the states themselves.  

The 1999 paper just cited obtained an upper bound on the 
cost of
the specific nonlocal measurement presented there, a bound that has recently been improved
and generalized by Cohen \cite{Cohen}.  In addition, 
there are in the literature at least three other avenues of research that bear on the problem of
finding the entanglement cost of nonlocal measurements.
First, there are several papers that simplify or extend the results of Ref.~\cite{nwoe}, for example by
finding other examples of measurements with product-state outcomes that cannot be
carried out locally \cite{UPB, DiVincenzo, DiVincenzo2, Groisman, WalgateHardy, Cohen,
Cohen1, Wootters, Duan, Feng, Koashi}.  A related line of research asks
whether or not a given set of orthogonal bipartite or multipartite states (not necessarily a complete basis, and not necessarily unentangled) can be distinguished by LOCC \cite{Chen, Chen2, Ghosh, Horodecki, Virmani, WalgateHardy, Walgate, Acin, Ogata, Eggeling, Chefles2, Bandyopadhyay, Nathanson, Owari, Fan, Watrous, Hayashi, Duan2, Chen3, WalgateScott}, and if not, how well one {\em can} distinguish the states by such means \cite{Badziag, Cao, Hillery, Horodecki2, Terhal}.  Finally, a number of authors have investigated the cost in entanglement, or the entanglement production capacity, of various bipartite and multipartite operations \cite{Chefles, Collins, Cirac, Berry, Dur, Eisert, Leifer, Ye, BennettHarrow, Kraus1, Zanardi, Groisman2, Huelga, Huelga2, Reznik2, Dur2, Reznik, Jozsa}.

In this paper we consider three specific cases:  (i) a class of orthogonal measurements on 
two qubits, in which the four eigenstates are equally entangled, (ii) a somewhat broader class of orthogonal measurements with
unequal entanglements, and (iii) a general, nonorthogonal, bipartite measurement in $d \times d$ dimensions
that is
invariant under all local Pauli operations.  
For the first of our three cases we present upper and lower
bounds on the entanglement cost.  For the second case we obtain a lower bound,
and for the last case we compute the cost exactly: it is equal to the average entanglement
of the states associated with the outcomes.   Throughout the paper, we mark our main results
as Propositions.

The upper bound in case (i) can be obtained directly from a protocol devised by Berry \cite{Berry}---a refinement of
earlier protocols \cite{Cirac, Ye}---for performing a closely related
nonlocal
unitary transformation.  Our bound is therefore the same as Berry's bound.  However, because we
are interested in performing a measurement rather than a unitary transformation, we give an alternative
protocol consisting of a sequence of local measurements.     

To get our lower bounds, we use a method developed
in papers on the local distinguishability of bipartite states \cite{Smolin, Ghosh, Horodecki}.
The average entanglement between
two parties cannot be increased by LOCC; so in performing the measurement, 
the participants must consume at least
as much entanglement as the measurement can produce.  This fact is the basis of all but one
of our lower bounds.  The one exception is in Section III, where we use a more stringent
condition, a bound on the success probability of local entanglement manipulation, to put a tighter bound on the cost for a limited class of procedures.

\subsection{1. Statement of the Problem}

To define the entanglement cost, we imagine two participants, Alice and Bob,
each holding one of the two objects to be measured.  We allow them to do any sequence
of local operations and classical communication, but we do not allow them to transmit
quantum particles from one location to the other.  Rather, we give them, as a resource,
arbitrary shared entangled states,
and we keep track of the amount of entanglement they consume in performing the measurement.   

At this point, though, we have a few options in defining the problem.  Do we try to find the cost of performing the 
measurement only once, or do we imagine that the same measurement will be performed
many times (on many different pairs of qubits) and look for the asymptotic cost per trial?
And how do we quantify the amount of entanglement that is used up?
In this paper we imagine that Alice and Bob will perform the given measurement only
once.  (In making this choice we are following Cohen \cite{Cohen}.)
However, we suppose that this measurement is one of many measurements they
will eventually perform (not necessarily repeating any one of the measurements and not necessarily
knowing in advance what the future measurements will be), and we assume that they have a large 
supply of entanglement from which they will continue to draw as they carry out these 
measurements.  In this setting it makes sense to use the standard measure of entanglement
for pure states, 
namely, the entropy of either of the two parts \cite{concentrate}.  Thus, for a pure state $|\psi\rangle$ 
of a bipartite system AB, the entanglement is
\begin{equation}
{\mathcal E}(|\psi\rangle) = -\hbox{tr} \rho_A \log \rho_A,   \label{basicE}
\end{equation}
where $\rho_A$ is the reduced density matrix of particle A: $\rho_A = \hbox{tr}_B |\psi\rangle \langle \psi |$.  In this paper, the logarithm will always be base two; so the entanglement is measured
in ebits.  
By means of local operations and classical communication, Alice and Bob can create from their
large supply of entanglement any specific state that they need.  For example, if they
create and completely use up a copy of the state $|\phi^+_c\rangle = c|00\rangle + d|11\rangle$,
this counts as a cost of ${\mathcal E}(|\phi^+_c\rangle) = -(c^2 \log c^2 + d^2 \log d^2)$.  
On the other hand, if their procedure converts an entangled state into a less entangled state, the cost
is the difference, that is, the amount of entanglement lost.  

A general measurement is
specified by a POVM, that is, a collection of positive 
semi-definite operators $\Pi_i$
that sum to the identity, each operator being associated with one of the outcomes of 
the measurement.  In this paper we restrict our attention to {\em complete} measurements,
that is, measurements in which each operator is of rank one; so each $\Pi_i$ is of the form
$\alpha_i|\phi_i\rangle\langle\phi_i|$ for some $\alpha_i$ in the range $0<\alpha_i\le1$.  In a complete {\em orthogonal} measurement, each operator is
a projection operator ($\alpha_i = 1$) that projects onto a single vector (an eigenvector $|\phi_i\rangle$ of the measurement).  Now, actually performing a 
measurement will always entail performing some operation on the measured system.  All that we
require of this operation is that Alice and Bob both end up with an accurate 
classical record of the outcome of the measurement.  In particular, we do not insist that
the measured system be collapsed into some particular state or even that it survive the measurement.  

We allow the possibility of probabilistic measurement procedures, in which the probabilities
might depend on the initial state of the system being measured.  However, we do not want our quantification of the
cost of a measurement to depend on this initial state; we are trying to characterize the 
measurement itself, not the system on which it is being performed.  So we assume that 
Alice and Bob are initially completely ignorant of the state of the particles they are measuring.
That is, the state they initially assign to these particles is the completely mixed state.  This is
the state we will use in computing any probabilities associated with the procedure.  

Bringing together the above considerations, we now give 
the definition of the quantity we are investigating in 
this paper.  Given a POVM $M$, let ${\mathcal P}(M)$ be 
the set of all LOCC procedures $P$ such that (i) $P$ uses 
pure entangled pairs, local operations, and classical communication, and (ii) $P$ realizes 
$M$ exactly, in the sense that for any initial state of the 
system to be measured, $P$ yields classical outcomes with 
probabilities that agree with the probabilities given by $M$. 
Then $C(M)$, the entanglement cost of a measurement $M$, is 
defined to be

\begin{equation}
C(M) = \inf_{P \in {\mathcal P}(M)} \left\langle 
{\mathcal E}_{\hbox{\scriptsize initial}} - {\mathcal D}_{\hbox{\scriptsize final}} \right\rangle,
\end{equation}
where ${\mathcal E}_{\hbox{\scriptsize initial}}$ is the total entanglement of all the 
resource states used in the procedure, ${\mathcal D}_{\hbox{\scriptsize final}}$ is
the distillable entanglement of the state remaining at the end of the procedure \cite{dist1,dist2}, 
and $\langle \cdots \rangle$ 
indicates an average over all the possible results of $P$, 
when the system on which the measurement is being performed 
is initially in the completely mixed state
 \footnote{One might wonder why we are using the {\em average} cost if we are imagining
each measurement being performed only once.  The reason is this: even in a series of distinct
measurements, if the series is long enough the actual cost will, with very high probability, be
very close to the sum of the average costs of the individual measurements.}.  (Though we allow
and take into account the possibility of some residual 
entanglement ${\mathcal D}_{\hbox{\scriptsize final}}$, in all the 
procedures we consider explicitly in this paper, the entanglement in the 
resource states will in fact be used up completely.) 

A different notion of the entanglement cost of a measurement is considered in Ref.~\cite{Jozsa}, namely,
the amount of entanglement needed to effect a Naimark extension of a given POVM\@.  In that case the
entanglement is between the system on which the POVM is to be performed and an ancillary system needed to make
the measurement orthogonal.  For any orthogonal measurement, and indeed for all the measurements considered
in this paper, the entanglement cost in the sense of Ref.~\cite{Jozsa} is zero.  

\subsection{2.  Measurements and unitary transformations}

One way to perform a nonlocal orthogonal measurement on a bipartite system is to perform a nonlocal
unitary transformation that takes the eigenstates of the desired measurement into 
the standard basis, so that the measurement can then be finished locally.  
(We will use this fact in Section II.)  
So one might wonder whether the problem we are investigating in this paper, at least for the case 
of orthogonal measurements, is equivalent to the problem of finding the cost of a nonlocal unitary
transformation.  A simple example shows that the two problems are distinct.  

Suppose that Alice holds two qubits, labeled A$'$ and A, and Bob holds a single qubit labeled B.
They want to perform an orthogonal measurement having the following eight eigenstates.
\begin{equation}
\begin{split}
(1/\sqrt{2})(|000\rangle + |011\rangle), \hspace{6mm}|100\rangle \\
(1/\sqrt{2})(|000\rangle - |011\rangle), \hspace{6mm}|101\rangle \\
(1/\sqrt{2})(|001\rangle + |010\rangle), \hspace{6mm}|110\rangle \\
(1/\sqrt{2})(|001\rangle - |010\rangle), \hspace{6mm}|111\rangle
\end{split}
\end{equation}
Here the order of the qubits in each ket is A$'$, A, B\@.  Alice and Bob can carry out this measurement
by the following protocol: Alice measures qubit A$'$ in the standard basis.  If she gets the outcome
$|1\rangle$, she and Bob can finish the measurement locally.  If, on the other hand, she gets the
outcome $|0\rangle$, she uses up one ebit to teleport the state of qubit A to Bob, who then 
finishes the measurement.  The average cost of this protocol is 1/2 ebit, because the probability
that Alice will need to use an entangled pair is 1/2.  

On the other hand, one can show that any unitary transformation that could change the above basis into
the standard basis would be able to create 1 ebit of entanglement and must therefore
consume at least 1 ebit.  So the cost of the measurement in this case is strictly
smaller than the cost of a corresponding unitary transformation.  

The crucial difference is
that when one does a unitary transformation, one can gain no information about the
system being transformed.  So there can be no averaging between easy cases and
hard cases.

\subsection{3. Two general bounds on the cost}

There are two general bounds on $C(M)$, an upper bound and a lower bound, that apply to all complete bipartite measurements. 
These bounds are expressed in the following two Propositions.

\medskip

\noindent {\bf Proposition 1.}  Let $M$ be a POVM on two objects A and B, having state spaces
of dimensions $d_A$ and $d_B$ respectively.  Then $C(M) \le \min\{\log d_A, \log d_B\}$.  

\medskip

\noindent {\em Proof.}  Let Alice and Bob share, as a resource, a maximally entangled state of two 
$d_A$-dimensional objects.  They can use this pair to teleport 
the state of A from Alice to Bob \cite{teleport},
who can then perform the measurement $M$ locally.  The entanglement of the resource
pair is $\log d_A$.  So $\log d_A$ ebits are sufficient to perform the measurement.  Similarly,
$\log d_B$ ebits would be sufficient to teleport the state of B to Alice.   So the cost of $M$ is no greater than 
$\min\{\log d_A, \log d_B\}$.  \hfill $\square$

\medskip

As we have mentioned, most of our lower bounds are obtained by considering the entanglement production capacity of our measurements.  Specifically, we imagine that in addition to particles A and B,
Alice and Bob hold, respectively, auxiliary particles C and D.  We consider an initial state of the
whole system such that the measurement $M$ on AB collapses CD into a possibly entangled state
\cite{Smolin, Ghosh, Horodecki}.
The average amount by which the measurement increases the entanglement between
Alice and Bob is then a lower bound on $C(M)$.  That is, 
\begin{equation}
\begin{split}
C(M) \ge &\; \hbox{(average final entanglement
of CD)} \\
&- \hbox{(initial entanglement between AC and BD)}.
\end{split}
\end{equation}
In the proof of the following proposition, the initial entanglement is zero.

\medskip

\noindent {\bf Proposition 2.}  Let $M$ be a bipartite POVM consisting of the operators $\alpha_i|\phi_i\rangle\langle\phi_i|$, where each $|\phi_i\rangle$ is a normalized state of particles
A and B, each of which has a $d$-dimensional state space.  Then $C(M)$ is at least as great as the
average entanglement $\langle {\mathcal E} \rangle$ of the states $|\phi_i\rangle$.  That is, 
\begin{equation}
C(M) \ge \langle {\mathcal E} \rangle \equiv \frac{1}{d^2}\sum_i \alpha_i {\mathcal E}(|\phi_i\rangle).  \label{ave}
\end{equation}

\medskip

\noindent {\em Proof.} Let the initial state of ABCD be
\begin{equation}
|\Psi\rangle = \frac{1}{d}\sum_{kl} |kk\rangle_{AC}|ll\rangle_{BD},  \label{maxent}
\end{equation}
a tensor product of two maximally entangled states.  
Note that the reduced density matrix of particles A and B is the completely mixed state,
in accordance with our definition of the problem.  
When the measurement yields the outcome $i$, its effect on 
$|\Psi\rangle$ can be expressed in the
form \cite{Kraus}
\begin{equation}
|\Psi\rangle\langle\Psi| \rightarrow  \sum_j \left( A_{ij}\otimes I_{CD} \right) |\Psi\rangle\langle\Psi|
( A^\dag_{ij}\otimes I_{CD} ),  \label{op}
\end{equation}
where $I_{CD}$ is the identity on CD, and the operators $A_{ij}$ act on the state space of particles A and B, telling us what happens to the system when the $i$th outcome occurs.  
The trace of the right-hand side of Eq.~(\ref{op}) is not unity but is
the probability of the $i$th outcome.  (Note that $A_{ij}$ may send 
states of AB to a different state space, including, for example, the state space of the system
in which the classical record of the outcome is to be stored.  The index $j$ is needed because
the final state of the system when outcome $i$ occurs could be a mixed state.)  
The operators $A_{ij}$ satisfy
the condition
\begin{equation}
\sum_j A^\dag_{ij}A_{ij} = \Pi_i = \alpha_i|\phi_i\rangle\langle \phi_i|.
\end{equation}
Applying the operation of Eq.~(\ref{op}) to the state of Eq.~(\ref{maxent}), and then
tracing out everything except particles C and D, one finds that these particles are left 
in the state
\begin{equation}
(|\phi_i\rangle\langle \phi_i|)^*,
\end{equation}
where the asterisk indicates complex conjugation in the standard basis.  
This conjugation does not affect the entanglement; so, when outcome $i$ occurs,
particles C and D are left in a state with entanglement ${\mathcal E}(|\phi_i\rangle)$.  The 
probability of this outcome is $\alpha_i/d^2$.  So the average entanglement of CD 
after the measurement has been performed is the 
quantity $\langle {\mathcal E}\rangle$ of Eq.~(\ref{ave}).  But the average entanglement
between Alice's and Bob's locations cannot have increased as long as Alice and Bob
were restricted to local operations and classical communication.  So in the process
of performing the measurement, Alice and Bob must have used up an amount of
entanglement equal to or exceeding $\langle {\mathcal E} \rangle$.  \hfill $\square$

\medskip

In the following three sections we improve these two bounds for a specific measurement
that we label $M_a$, an orthogonal measurement on two qubits with eigenstates
given by
\begin{equation}
\begin{split}
&|\phi^+_a\rangle = a|00\rangle + b|11\rangle, \hspace{5mm} |\phi^-_a\rangle = b|00\rangle
-a|11\rangle   \\
&|\psi^+_a\rangle = a|01\rangle + b|10\rangle, \hspace{5mm} |\psi^-_a\rangle = b|01\rangle
-a|10\rangle \label{Mstates}
\end{split}
\end{equation}
Here $a$ and $b$ are nonnegative real numbers with $a \ge b$ and \hbox{$a^2 + b^2 = 1$}.  
Section II presents an improved upper bound for this measurement, Section III derives a lower bound for
a restricted class of procedures, and Section IV derives an absolute lower bound.  
We then consider a somewhat more general measurement in Section V.  

In Section VI we exhibit a class of bipartite measurements, in dimension $d\times d$,
for which we can find a procedure that achieves the lower bound of Eq.~(\ref{ave}).  As noted
earlier, these are the POVMs that are invariant under all local Pauli operations.


\section{II.  Upper bound for {\boldmath $M$}\hspace{-0.5mm}$_a$}

One way to perform the measurement $M_a$ is to perform the following unitary transformation
on the two qubits.  
\begin{equation}
U = e^{i\alpha \sigma_y \otimes \sigma_x} = \mtx{cccc}{a & 0 & 0 & b \\ 0 & a & b & 0 \\
0 & -b & a & 0 \\ -b & 0 & 0 & a}, \label{unitary}
\end{equation}
where $\cos\alpha = a$ and $\sin\alpha = b$, the matrix
is written in the standard basis and 
the $\sigma$'s are the usual Pauli matrices, 
\begin{equation} 
\sigma_x = \mtx{cc}{0 & 1 \\ 1 & 0} \hspace{2mm}\hbox{and}\hspace{2mm}
\sigma_y = \mtx{cc}{0 & -i \\ i & 0}.
\end{equation}
Under this transformation, the four orthogonal states
that define the measurement $M_a$ are transformed into
\begin{equation}
\begin{split}
\hfill&|\phi^+_a\rangle = a|00\rangle + b|11\rangle \rightarrow |00\rangle \hfill  \\
\hfill&|\phi^-_a\rangle = b|00\rangle - a|11\rangle \rightarrow -|11\rangle \hfill  \\
\hfill&|\psi^+_a\rangle = a|01\rangle + b|10\rangle \rightarrow |01\rangle \hfill \\
\hfill&|\psi^-_a\rangle = b|01\rangle - a|10\rangle \rightarrow -|10\rangle \hfill  
\end{split}
\end{equation}
So once the transformation has been done, the measurement $M_a$ can be completed 
locally; Alice and Bob both make the measurement $|0\rangle$ versus $|1\rangle$
and tell each other their results.  

The transformation $U$ is equivalent to one that has been analyzed in Refs.~\cite{Groisman2, Dur2, Cirac, Ye, Berry},
all of which give procedures that are consistent with the rules we have
set up for our problem; that is, the procedures can be used to perform the measurement
once, rather than asymptotically, using arbitrary entangled states as resources.  (Some
of those papers consider the asymptotic problem, but their procedures also work in the
setting we have adopted here.)  
It appears that the procedure presented by Berry in Ref.~\cite{Berry} is the most efficient
one known so far.  It is a multi-stage procedure, involving at each stage a measurement 
that determines whether another stage, and another entangled pair, are needed.  

We now present a measurement-based protocol for performing $M_a$.  The protocol can be derived from 
Berry's and yields the same upper bound on the cost, but we arrive at it 
in a different way that may have conceptual value in the analysis of other
nonlocal measurements. 


The construction of the protocol begins with the following observations.  If Alice were to try to 
teleport her qubit to Bob using as a resource an incompletely entangled pair, she would cause 
a nonunitary distortion in its state.  With his qubit and Alice's distorted qubit, Bob could, with some probability less than one, successfully complete the measurement.  However, if he gets the wrong
outcome, he will destroy the information necessary to complete the measurement.  We require
the measurement always to be completed, so this protocol fails.  On the other hand, suppose Alice, again using a partially entangled pair, performs an {\em incomplete} teleportation, conveying to Bob only one rather than two classical bits, and suppose Bob similarly makes an
incomplete measurement, extracting only one classical bit from his two qubits.  In that case, if the incomplete measurements
are chosen judiciously, a failure does not render the desired measurement impossible but only 
requires that Alice and Bob do a different nonlocal measurement on the qubits they now
hold.  In the following description of the
protocol, we have incorporated the unitary transformations associated with teleportation
into the measurements themselves, so that the whole procedure is a sequence of local
projective measurements.

Like Berry's protocol, our protocol consists a series of rounds, beginning with what we will call
``round one''.  
\begin{enumerate}
\item Alice and Bob are given as a resource the entangled state 
$|\phi^+_x\rangle = x|00\rangle + y|11\rangle$, where 
the positive real numbers $x$ and $y$ (with $x^2 + y^2 = 1$) are to be determined by minimizing the eventual cost.  Thus each participant holds two qubits: the qubit to be measured and 
a qubit that is part of the shared resource.  
\item Alice makes a binary measurement on her two qubits, defined by two orthogonal 
projection operators: 
\begin{equation}
\begin{split}
\hfill&P = |\Phi^+\rangle\langle \Phi^+| + |\Psi^-\rangle\langle \Psi^-|  \\
\hfill&Q = |\Phi^-\rangle\langle \Phi^-| + |\Psi^+\rangle\langle \Psi^+|  \label{preBell}
\end{split}
\end{equation}
Here the Bell states $|\Phi^\pm\rangle$ and $|\Psi^\pm\rangle$ are defined by 
$|\Phi^\pm\rangle
= (|00\rangle \pm |11\rangle)/\sqrt{2}$ and $|\Psi^\pm\rangle
= (|01\rangle \pm |10\rangle)/\sqrt{2}$.
Alice transmits (classically) the result of her measurement to Bob.
(Here Alice is doing the incomplete teleportation.  In a complete teleportation she would 
also distinguish $|\Phi^+\rangle$ from $|\Psi^-\rangle$, and  $|\Phi^-\rangle$ from $|\Psi^+\rangle$.)
\item If Alice gets the outcome $P$, Bob performs the following binary measurement on his 
two qubits:  
\begin{equation}
\begin{split}
&P_1 = |\phi_1^+\rangle\langle \phi_1^+| + |\psi_1^+\rangle\langle \psi_1^+|,  \\
&Q_1 = |\phi_1^-\rangle\langle \phi_1^-| + |\psi_1^-\rangle\langle \psi_1^-|.  
\end{split}
\end{equation}
Here $|\phi_1^+\rangle
= A|00\rangle + B|11\rangle$, $|\phi_1^-\rangle
= B|00\rangle - A|11\rangle$, $|\psi_1^+\rangle
= B|01\rangle + A|10\rangle$, and $|\psi_1^-\rangle
= A|01\rangle - B|10\rangle$, and the real coefficients $A$ and $B$ are obtained from $(a,b)$
and $(x,y)$ via the equation $Ax/a = By/b$, together with the normalization condition 
\hbox{$A^2+B^2=1$}.
(These values are chosen so as to undo the distortion caused by Alice's imperfect
teleportation.)
On the other hand, if Alice gets the outcome $Q$, Bob performs a different binary measurement:
\begin{equation}
\begin{split}
&P_2 = |\phi_2^+\rangle\langle \phi_2^+| + |\psi_2^+\rangle\langle \psi_2^+|, \\ 
&Q_2 = |\phi_2^-\rangle\langle \phi_2^-| + |\psi_2^-\rangle\langle \psi_2^-|.  
\end{split}
\end{equation}
Here $|\phi_2^+\rangle
= B|00\rangle + A|11\rangle$, $|\phi_2^-\rangle
= A|00\rangle - B|11\rangle$, $|\psi_2^+\rangle
= A|01\rangle + B|10\rangle$, and $|\psi_2^-\rangle
= B|01\rangle - A|10\rangle$.  
\item If Alice and Bob have obtained either of the outcomes $P\otimes P_1$ or $Q \otimes Q_2$, 
which we call the ``good'' outcomes, they
can now finish the desired measurement $M_a$ by making local measurements, with no further
expenditure of entangled resources.  For example, if they get the outcome 
$P\otimes P_1$, Alice now distinguishes between $|\Phi^+\rangle$ and $|\Psi^-\rangle$
(which span the subspace picked out by $P$), and Bob distinguishes between
$|\phi_1^+\rangle$ and $|\psi_1^+\rangle$ (which span the subspace picked out
by $P_1$). 
The total probability of getting one of the two good outcomes
is
\begin{equation} \label{probability}
\hbox{probability}\; = \frac{1}{(a/x)^2 + (b/y)^2}.
\end{equation}
On the other hand, if they have obtained one of the other
two outcomes, $P\otimes Q_1$ or $Q \otimes P_2$---the ``bad'' outcomes---they find that in order to finish the measurement
$M_a$ on their {\em original} pair of qubits, they now have to perform a different 
measurement $M_{a_2}$
on the system that they now hold.  (Even though each participant started with two qubits, each of them has now distinguished a pair of two-dimensional subspaces, effectively removing one qubit's worth of quantum information.  So the remaining quantum information
on each side can be held in a single qubit.)  The measurement $M_{a_2}$ has the same form as
$M_a$, but with new values $a_2$ and $b_2$ instead of $a$ and $b$.  
The new values are determined by the equations
\begin{equation} \label{newa}
a_2=\frac{(x^2 - y^2)ab}{\sqrt{x^4b^2 + y^4a^2}} \hspace{6mm} b_2 = \sqrt{1-a_2^2}\, .
\end{equation}
In any case, Alice and Bob have now finished round one.  If they have obtained one of the
bad outcomes, 
they now have
two choices: (i) begin 
again at step 1 but with the new values $a_2$ and $b_2$, or (ii) use up a whole ebit to teleport Alice's system to Bob, who finishes the measurement locally.  They choose the method that will ultimately 
be less costly in entanglement.  If they choose option (i), we say that they have begun round two.  
\item This procedure is iterated until the measurement is finished or until $L$ rounds have been
completed, where $L$ is an integer chosen in advance.  In round $j$, the measurement parameter
$a_j$ is determined from the parameters $a_{j-1}$ and $x_{j-1}$ used in the preceding round
according to Eq.~(\ref{newa}) (with the appropriate substitutions).  Here $a_1$ and $x_1$
are to be interpreted as the first-round values $a$ and $x$.  
\item If $L$ rounds are completed and the measurement is still unfinished, Alice teleports
her system to Bob, who finishes the measurement locally.  
\end{enumerate}

The entanglement used in stage $j$ of this procedure is 
${\mathcal E}(|\phi^+_{x_j}\rangle)=h(x_j^2)$, where $h$ is the binary entropy 
function $h(z) = -[z\log z + (1-z)\log(1-z)]$.  From Eqs.~(\ref{probability}) 
and (\ref{newa}),
we therefore have the following upper bound on the cost of the measurement $M_a$.

\medskip

\noindent {\bf Proposition 3.} For each positive integer $j$, let $x_j$ satisfy $0 < x_j < 1$.  We define the functions $F(a,x)$ (failure probability) and 
$a'(a,x)$ (new value of the measurement parameter) as follows:
\begin{equation}
\begin{split}
& F(a,x) = 1 - \frac{1}{(a/x)^2 + (b/y)^2} \\
& a'(a,x) = \frac{(x^2 - y^2)ab}{\sqrt{x^4b^2 + y^4a^2}},
\end{split}
\end{equation}
where $y = (1-x^2)^{1/2}$ and $b= (1-a^2)^{1/2}$.  Let
$B_1(a;x) = h(x^2) + F(a,x)$, and for each integer $n \ge 2$, let $B_n(a;x_1,\ldots, x_n)$
be defined by 
\begin{equation}
\begin{split}
&B_n(a;x_1,\ldots, x_n)
= h(x_1^2) \\
&+ F(a,x_1)B_{n-1}[a'(a,x_1);x_2, \ldots, x_n].
\end{split}
\end{equation}
Then for each positive integer $n$, $B_{n}(a; x_1, \cdots, x_{n})$ is an upper bound on $C(M_a)$.

\medskip

The protocol calls for minimizing the bound over the values of $n$ and $x_j$.  This optimization 
problem is exactly the problem analyzed
by Berry.  We present in Fig.~1 the minimal cost as obtained by a numerical optimization, plotted as a function
of the entanglement of the eigenstates of the measurement.  (In constructing the curve,
we have limited Alice and Bob to two rounds.  Additional rounds do not make 
a noticeable difference in the shape of the curve, given our choice of the axis
variables.)  We also show on the figure the lower
bound to be derived in Section IV.  We note that so far, for cases in which the entanglement
of the eigenstates of $M_a$ exceeds around 0.55 ebits, there is no known
measurement strategy that does better than simple teleportation, with a cost of one ebit.  
\begin{figure}[h] 
\begin{tabular}{l}
Bounds on the cost \\
\includegraphics[scale=1]{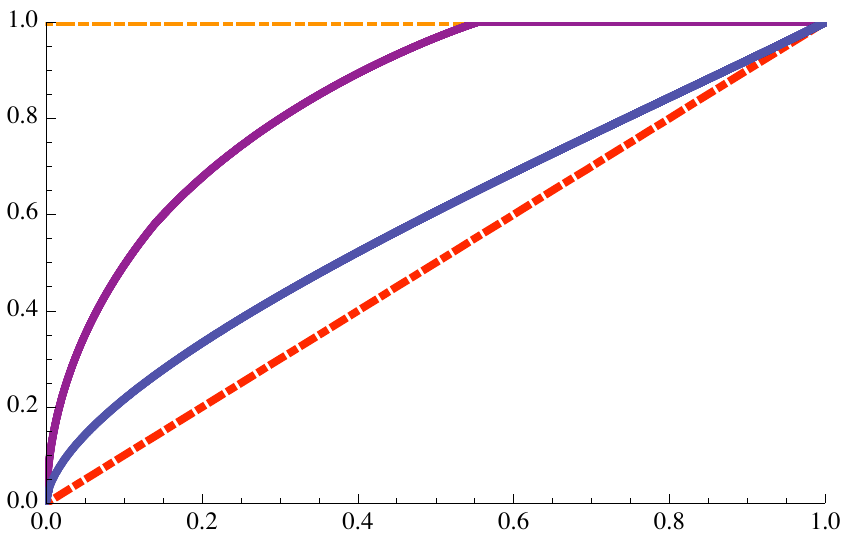}
\end{tabular}

{\centering Entanglement of the states}
\smallskip
\caption{The solid curves are upper and lower bounds on the entanglement cost of the measurement $M_a$.  (The derivation of the lower bound is in Section IV.)  The
diagonal dashed line is the general lower bound defined by the entanglement of the states
themselves, and the horizontal dashed line is the general upper bound based on teleportation.} 
\end{figure}

\section{III. Limitation to a single round}

As it happens, most of the savings in the above strategy---compared to the cost of simple teleportation---already
appears in the first round.  We now 
consider the single-round case in more detail.  It turns out that, at least for small
values of the entanglement of the eigenstates, we can determine quite precisely the
minimal cost of the measurement $M_a$ when Alice and Bob are 
restricted to a single round.

We begin by defining the class of measurement strategies we consider in this section.  
A ``single-round procedure'' is a measurement procedure
of the following form.  (i) Alice and Bob are given the state $x|00\rangle+y|11\rangle$ at first, with 
which they try to complete the measurement.  (ii) If they use this resource but fail
to carry out the measurement, Alice teleports a qubit to Bob, who finishes the measurement
locally.  (For the procedure outlined in the preceding section,
this restriction amounts to setting $L$ equal to 1.)  We refer to the minimum entanglement cost entailed by any such procedure
as the ``single-round cost".  In this section we find upper and lower bounds on the
single-round cost of $M_a$.  

The minimal cost of the specific procedure outlined in Section II, when it is restricted to a single round,
is given by 
\begin{equation} \label{oneround}
\hbox{cost}\; = h(x^2) +\left[1-  \frac{1}{(a/x)^2 + (b/y)^2}\right],
\end{equation}
where the value 
of $x$ is chosen so as to minimize the cost.  (Here $x^2 + y^2 = 1$ as before.)  
The two terms of Eq.~(\ref{oneround}) are easy to interpret: the first term is the entanglement
of the shared resource that is used up in any case, and the second term, obtained 
from Eq.~(\ref{probability}), is the probability
of failure (multiplied by the 1 ebit associated with the resulting teleportation).  
Numerically minimizing the cost over values of $x$, we obtain the upper curve
in Fig.~2, which is thus an upper bound on the single-round cost of $M_a$.   
The same upper bound was obtained by Ye, Zhang, and Guo for performing the 
corresponding nonlocal unitary transformation \cite{Ye}.

We can also find a good lower bound for such procedures, using
a known upper bound on the probability of achieving a certain increase in the entanglement of 
a single copy 
through local operations and classical communication \cite{Vidal, Jonathan}.   As in all our lower
bound arguments, we consider a state in which qubits A and B are initially entangled with auxiliary qubits C and D,
which will not be involved in the measurement.
(As before, Alice holds qubits A and C,
and Bob holds B and D.)  For our present purpose, we choose the initial state to be
\begin{equation}  \label{fourqubits}
\begin{split}
|\xi\rangle = \frac{1}{2}&\left[{|\phi^+_a\rangle}_{AB} {|\phi^+_c\rangle}_{CD}
+ {|\phi^-_a\rangle}_{AB} {|\phi^-_c\rangle}_{CD} \right.  \\
&+\left.  {|\psi^+_a\rangle}_{AB} {|\psi^+_c\rangle}_{CD}
+ {|\psi^-_a\rangle}_{AB} {|\psi^-_c\rangle}_{CD} \right].
\end{split}
\end{equation}
Here the states with the index $c$ are defined as in Eq.~(\ref{Mstates}), but with 
$c$ and $d$ in place of $a$ and $b$.  
We assume for definiteness
that $1 > c > d > 0$, $c$ and $d$ to be determined later.
Note that again the reduced state of qubits A and B, after tracing out the auxiliary qubits,
is the completely mixed state, as it must be to be consistent with our 
definition of the entanglement cost.    
One can show directly from Eq.~(\ref{fourqubits}) that the eigenvalues
of the density matrix of Alice's (or Bob's) part of the system, that is, the squared Schmidt
coefficients, are
\begin{equation}
(ac+bd)^2  \hspace{2mm} \hbox{and} \hspace{2mm} (ad-bc)^2.
\end{equation}
In addition to these qubits, Alice and Bob hold their entangled resource,
which we can take without loss of generality to be in the state
\begin{equation}
|\phi^+_x\rangle = x|00\rangle + y|11\rangle.
\end{equation}
They now try to execute the measurement by using up this resource.

If Alice and Bob succeed in distinguishing
the four states $\{|\phi^+_a\rangle, |\phi^-_a\rangle, |\psi^+_a\rangle,
|\psi^-_a\rangle\}$, they will have collapsed qubits C and D into one of the
four corresponding states represented in $|\xi\rangle$.  Each of these states
has Schmidt coefficients $c^2$ and $d^2$.  Using a result of 
Jonathan and Plenio \cite{Jonathan}, we can place an upper bound on the probability
of achieving the transformation from the state $|\xi\rangle \otimes |\phi^+_x\rangle$
to one of the four desired final states of qubits C and D.  This probability cannot be larger
than 
\begin{equation}  \label{JP}
\frac{\sum_{j=\ell}^4 \alpha_j}{\sum_{j=\ell}^2 \beta_j},
\end{equation}
where $\alpha_j$ and $\beta_j$ are, respectively, the squared Schmidt coefficients
of the initial state and any of the desired final states, in decreasing order.  
(There are at most four nonzero Schmidt coefficients in the initial state; hence the
upper limit 4 in the numerator.  Similarly, the upper limit 2 in the denominator
reflects the fact that the final state, a state of C and D, has at most
two nonzero Schmidt coefficients.)  In general $\ell$
can take any value from 1 to the number of nonzero Schmidt coefficients
of the final state.  In our problem there are only two values of $\ell$ to consider.  
The case $\ell=1$ tells us only that the probability does not exceed unity; so 
the only actual constraint comes from the case $\ell=2$, which tells us that
\begin{equation}
\hbox{the success probability} \; \le \frac{1 - (ac+bd)^2 x^2}{1-c^2}.
\end{equation}
The cost of any single-round procedure is therefore at least
\begin{equation}  \label{concave}
\hbox{cost}\; \ge h(x^2) + \max\left\{ 0, \left[1 - \frac{1 - (ac+bd)^2 x^2}{1-c^2}\right]\right\},
\end{equation}
since a failure will lead to a cost of one ebit for the teleportation.  

\begin{figure}[h]
\begin{tabular}{l}
Bounds on the single-round cost\\
\includegraphics[scale=1]{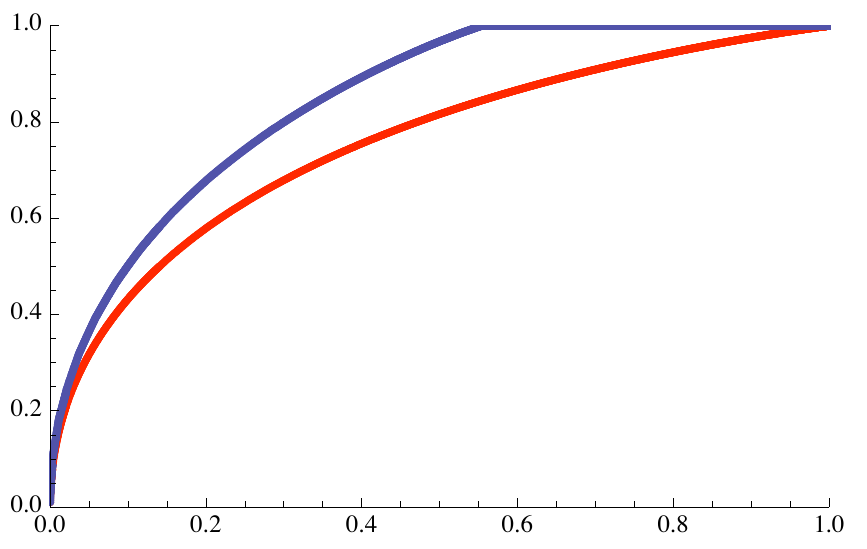}
\end{tabular}

{\centering Entanglement of the states}
\smallskip
\caption{Upper and lower bounds on the entanglement cost of the measurement $M_a$, when
Alice and Bob are restricted to a single round before resorting to teleportation.} \label{oneroundgraph}
\end{figure}

Alice and Bob will choose their resource pair, that is, they will choose
the value of $x$, so as to minimize the cost.  So we want to find a value
of $x$ that minimizes the right-hand side of Eq.~(\ref{concave}).  Because the 
probability of failure cannot be less than zero, we can restrict our attention
to values of $x$ in the range
\begin{equation}  \label{constraint}
c/(ac+bd) \le x \le 1.
\end{equation}
In this range, the cost is a concave function of $x^2$; so the function
achieves its minimum value at one of the two endpoints.  
We thus have the following lower bound on the cost
of any single-round procedure:
\begin{equation}  \label{ABchoice}
\hbox{cost}\; \ge \min\left\{ h\left[\frac{c^2}{(ac+bd)^2}\right] ,
\frac{(ac+bd)^2 - c^2}{1-c^2} \right\}.
\end{equation}

This bound holds for any value of $c$ for which it is defined.  To make the bound as strong
as possible, we want to maximize it over all values of $c$.  In the range
$1/\sqrt{2} \le c \le a$, the 
first entry in Eq.~(\ref{ABchoice}) is a decreasing function of $c$,
whereas the second entry is increasing.  (For larger values of $c$,
both functions are decreasing until they become undefined
at $c = (ac+bd)$.  Beyond this point we would violate Eq.~(\ref{constraint}).)
Therefore, we achieve the strongest bound when the two entries are
equal.  That is, we have obtained the following result.

\medskip

\noindent {\bf Proposition 4.}  The single-round cost of the measurement $M_a$ is
bounded below by the quantity
\begin{equation}
\frac{(ac+bd)^2 - c^2}{1-c^2},    \label{lowerbound1}
\end{equation}
where $d = (1-c^2)^{1/2}$ and $c$ is determined by the equation
\begin{equation}
h\left[\frac{c^2}{(ac+bd)^2}\right]=\frac{(ac+bd)^2 - c^2}{1-c^2}.  \label{lowerbound2}
\end{equation}

\medskip

We have solved this equation numerically for a range of values of 
$a$ and have obtained the lower of the two curves in Fig.~2.  
For very weakly entangled eigenstates---that is, at the left-hand end
of the graph where the parameter $b$ is small---the single-round upper bound and the single-round
lower bound shown in the figure are very close to each other.  In fact, we find analytically that
for small $b$, both the upper and lower bounds
can be approximated by the function $2b\sqrt{\log (1/b)}$, in the sense that the ratio of each bound
with this function approaches unity as $b$ approaches zero.  (See the Appendix for
the argument.)  Or, in terms of the entanglement ${\mathcal E}$ of the states, we can say that for small $b$,
the single-round cost of the measurement is approximately equal to 
$\sqrt{2{\mathcal E}}$.
Thus in this limit, we have a very
good estimate of the cost of the measurement, but only if we restrict
Alice and Bob to a single round.  
We would prefer to have a lower bound that applies to any conceivable
procedure, and that is still better than the general
lower bound we derived in Section I.  We obtain such a bound in the following section.

\section{IV.  An absolute lower bound for {\boldmath $M$}\hspace{-0.5mm}$_a$}

Again, we imagine a situation in which Alice and Bob hold two auxiliary qubits C and D
that will not be involved in the measurement.  We assume the same initial state
as in Section III:
\begin{equation}
\begin{split}
|\xi\rangle = \frac{1}{2}&\left[{|\phi^+_a\rangle}_{AB} {|\phi^+_c\rangle}_{CD}
+ {|\phi^-_a\rangle}_{AB} {|\phi^-_c\rangle}_{CD} \right.  \\
&+\left.  {|\psi^+_a\rangle}_{AB} {|\psi^+_c\rangle}_{CD}
+ {|\psi^-_a\rangle}_{AB} {|\psi^-_c\rangle}_{CD} \right].
\end{split}
\end{equation}
As before, we are interested in the entanglement between Alice's part of the system
and Bob's part, that is, between AC and BD\@.  This entanglement is
\begin{equation}
{\mathcal E}_{\hbox{\scriptsize initial}} = h[(ac+bd)^2].
\end{equation}
If Alice and Bob perform
the measurement $M_a$, the final entanglement of CD is
\begin{equation}
{\mathcal E}_{\hbox{\scriptsize final}} = h(c^2).
\end{equation}
The quantity ${\mathcal E}_{\hbox{\scriptsize final}} - {\mathcal E}_{\hbox{\scriptsize initial}}$
is thus a lower bound on the cost of $M_a$, as expressed in the following proposition. 

\medskip

\noindent {\bf Proposition 5.}  Let $c$ satisfy $0 \le c \le 1$, and let $d=(1-c^2)^{1/2}$.  
Then $C(M_a) \ge h(c^2) - h[(ac+bd)^2]$.

\medskip

By maximizing this quantity numerically over the parameter $c$, we get our best absolute lower
bound on the entanglement cost $C(M_a)$.  This bound is plotted in 
Fig.~1.  
What is most interesting about this bound is that, except at the extreme points
where the eigenstates of the measurement are either all unentangled or
all maximally entangled, the bound is strictly larger than the entanglement
of the eigenstates themselves.  This is another example, then, showing
that the nonseparability of the measurement can exceed the nonseparability 
of the states that the measurement distinguishes.

Not only is our new lower bound absolute in the sense that it does not depend 
on the number of rounds used by Alice and Bob; it applies even asymptotically. 
Suppose, for example, that Alice and Bob are given $n$ pairs of qubits and 
are asked to perform the same measurement $M_a$ on each pair.  It is 
conceivable that by using operations that involve all $n$ pairs, Alice and 
Bob might achieve an efficiency not possible when they are performing
the measurement only once.  Even in this setting, the lower bound given
in Proposition 5 applies.  That is, the cost of performing the measurement
$n$ times must be at least $n$ times our single-copy lower bound.  To see
this, imagine that {\em each} of the given pairs of qubits is initially entangled
with a pair of auxiliary qubits.  Both the initial entanglement of the whole system
(that is, the entanglement between Alice's side and Bob's side), and the 
final entanglement after the measurement, are simply proportional to $n$,
so that the original argument carries over to this case.

It is interesting to look at the behavior of the upper and lower bounds as 
the parameter
$b$ approaches zero, that is, as the eigenstates of the measurement approach
product states.  Berry has done this analysis for the upper bound and has found
that for small $b$, the cost is proportional to $b$, with proportionality constant
5.6418.  For our lower bound, it is a question of finding the value of $c$
(with $c^2+d^2 = 1$) that maximizes the difference
\begin{equation}
h(c^2) - h[(ac+bd)^2] \label{diff}
\end{equation}
for small $b$.
One finds that for small $b$, the optimal value of $c$ approaches the constant 
value $c = 0.28848$ (the numerical solution to the equation $(d^2 - c^2)\ln(d/c) = 1$), for 
which the bound 
is approximately equal to $1.9123b$.  Comparing this result
with the upper bound in the limit of vanishingly small entanglement, $5.6418b$, we see that there is still a sizable
gap between the two bounds.

The same limiting form, $1.9123b$, appears in Ref.~\cite{Dur} as the entanglement production
capacity of the unitary transformation of Eq.~(\ref{unitary}) for small $b$.  In fact, by extending the argument of Ref.~\cite{Dur} to non-infinitesimal
transformations, one obtains the entire lower-bound curve in Fig.~1.  Thus our lower bound for the cost
of the measurement $M_a$ is also a lower bound for the cost of the corresponding unitary
transformation.  We note, though, that the two optimization problems are not quite the same.  
To get a bound on the cost of the measurement, we maximized
$h(c^2) - h[(ac+bd)^2]$.  To find the entanglement production capacity of the unitary transformation,
one maximizes $h[(ac+bd)^2]-h(c^2)$.  Though the questions are different, it is not hard to show that maximum {\em value} is the same in both
cases. 

\section{V. Eigenstates with Unequal Entanglements}

We now consider the following variation on the measurement $M_a$. 
It is an orthogonal measurement, which we call $M_{a,c}$, with eigenstates
\begin{equation}
\begin{split}
&|\phi^+_a\rangle = a|00\rangle + b|11\rangle, \hspace{5mm} |\phi^-_a\rangle = b|00\rangle
-a|11\rangle,   \\
&|\psi^+_c\rangle = c|01\rangle + d|10\rangle, \hspace{5mm} |\psi^-_c\rangle = d|01\rangle
-c|10\rangle,
\end{split}
\end{equation}
where all the coefficients are real and nonnegative 
and all the states are normalized.  For this measurement
we again use the entanglement production argument to get a lower bound.
In this case we take the initial state of qubits ABCD to be
\begin{equation}
\begin{split}
|\eta\rangle = \frac{1}{2}&\left[|\phi^+_a\rangle_{AB}|\phi^+_{a'}\rangle_{CD}
+|\phi^-_a\rangle_{AB}|\phi^-_{a'}\rangle_{CD} \right.  \\
&\left. + |\psi^+_c\rangle_{AB}|\psi^+_{c'}\rangle_{CD}
+|\psi^-_c\rangle_{AB}|\psi^-_{c'}\rangle_{CD}\right],  \label{genstate}
\end{split}
\end{equation}
where the real parameters $a'$ and $c'$ are to be adjusted to achieve the
most stringent lower bound.  This initial state has an entanglement
between Alice's location and Bob's location (that is, between AC and BD) equal to 
the Shannon entropy of the following four probabilities:
\begin{equation}
\begin{split}
(aa' + bb' + cc' + dd')^2/4, \hspace{5mm} (aa' + bb' - cc' - dd')^2/4, \\
(ab' - ba' + dc' - cd')^2/4, \hspace{5mm} (ab' - ba' -dc' + cd')^2/4. 
\end{split}
\end{equation}
Once the measurement is completed, the 
final entanglement of the CD system, on average,
is
\begin{equation}
[h(a'^2) + h(c'^2)]/2.
\end{equation}
The difference between the final entanglement and the initial entanglement
is a lower bound on $C(M_{a,c})$, which we want to maximize by our choice 
of $a'$ and $c'$.  We have again done the maximization numerically, for many values of
the measurement parameters $a$ and $c$, covering their domain quite densely.  
We plot the results in Fig.~3.  
\begin{figure}[h]
\begin{tabular}{l}
Lower bound on the cost\\
\includegraphics[scale=0.76]{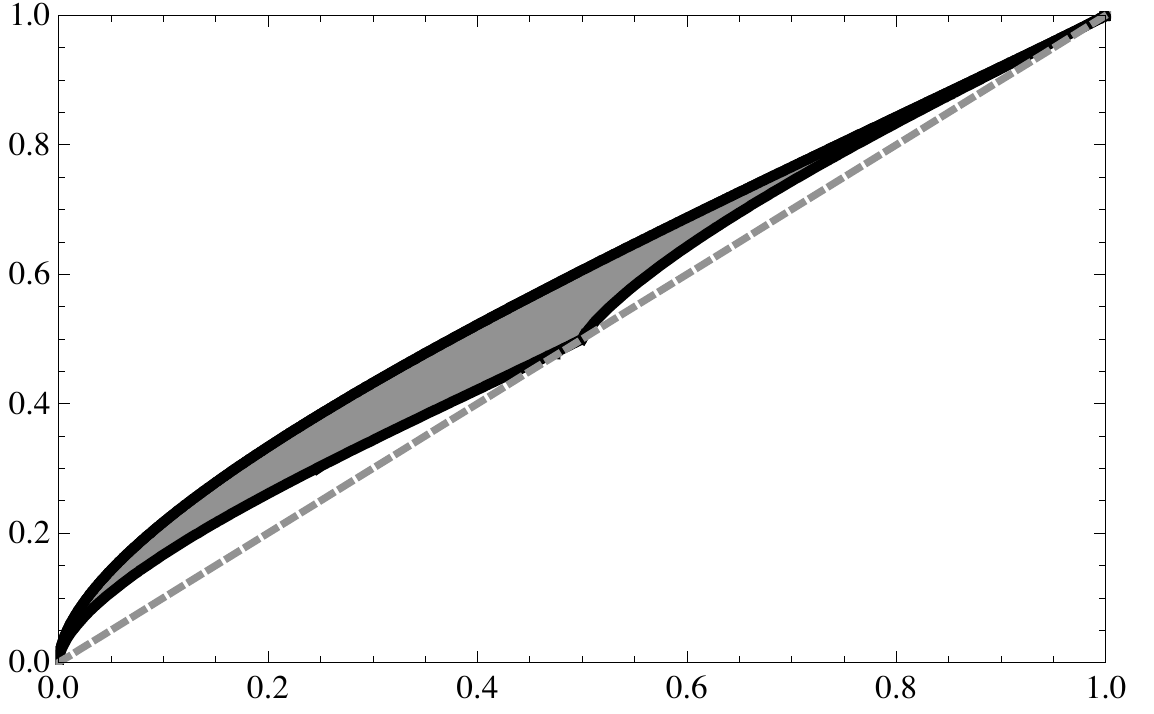}
\end{tabular}

{\centering Average entanglement of the states}
\smallskip
\caption{Lower bound for the measurement $M_{a,c}$ as computed from the pure state given in Eq.~(\ref{genstate}), plotted against the average entanglement
of the eigenstates (dashed line).  Different values of the pair $(a,c)$
can be associated with the same point on the horizontal axis but may yield
different lower bounds, as indicated by the gray area.  The point touching the
dashed line in the middle of the graph represents a measurement with two Bell
states and two product states, as in Eq.~(\ref{M3}).} 
\end{figure}
In almost every case, the resulting lower bound is {\em higher} than the average
entanglement of the eigenstates of the measurement.  The only exceptions
we have found, besides the ones already mentioned in Section IV (in which all the 
states are maximally entangled or all are unentangled), are those for which two of the measurement eigenstates
are maximally entangled and the other two are unentangled.  That is, this method
does not produce a better lower bound for the measurement with eigenstates
\begin{equation}
|\Phi^+\rangle, \; |\Phi^-\rangle, \; |01\rangle, \; |10\rangle, \label{M3}
\end{equation}
or for the analogous measurement with $|\Phi^\pm\rangle$ replaced
by $|\Psi^\pm\rangle$ and with the product states suitably replaced to 
make the states mutually orthogonal.  In all other cases the cost of the measurement is strictly
greater than the average entanglement of the states.

The measurement $M_{a,c}$ has been considered in Ref.~\cite{Ghosh}, whose results
likewise give a lower bound on the cost: $C(M_{a,b}) \ge 1-\log(a^2 + c^2)$ (where $a \ge b$
and $c \ge d$).  This bound is weaker than the one we have obtained, in part because
we have followed the later paper Ref.~\cite{Horodecki} in assuming an initial pure state rather than
a mixed state of ABCD.

\section{VI.  Measurements for Which the General Lower Bound Can Be Achieved}

Here we consider a class of measurements for which the cost {\em equals} the average
entanglement of the states associated with the POVM elements.  We begin with another two-qubit
measurement, which we then generalize to arbitrary dimension.  

\subsection{1.  An eight-outcome measurement}

A measurement closely related to $M_a$ is measurement $M^{(8)}_{a}$, which has eight outcomes,
represented by a POVM 
whose elements $\Pi_i = \alpha_i|\phi_i\rangle\langle\phi_i|$
all have $\alpha_i = 1/2$, with the eight states $|\phi_i\rangle$ given by
\begin{equation}
\begin{split}
|\phi^+_a\rangle = a|00\rangle + b|11\rangle, \hspace{5.2mm}& |\phi^-_a\rangle = b|00\rangle
-a|11\rangle \phantom{;}  \\
|\psi^+_a\rangle = a|01\rangle + b|10\rangle, \hspace{5mm}& |\psi^-_a\rangle = b|01\rangle
-a|10\rangle \phantom{.}  \\
|\phi^+_b\rangle = b|00\rangle + a|11\rangle, \hspace{5.2mm}& |\phi^-_b\rangle = a|00\rangle
-b|11\rangle  \phantom{;}  \\
|\psi^+_b\rangle = b|01\rangle + a|10\rangle, \hspace{5mm}& |\psi^-_b\rangle = a|01\rangle
-b|10\rangle. 
\end{split}
\end{equation}
That is, they are the same states as in $M_a$, plus the four states obtained by
interchanging $a$ and $b$.  Thus, Alice and Bob could perform the measurement
$M^{(8)}_a$ by flipping a fair coin to decide whether to perform $M_a$ or $M_b$.  
This procedure yields the eight possible outcomes: there are two possible outcomes of the coin toss,
and for each one, there are four possible outcomes of the chosen measurement.
The 
coin toss requires no entanglement; so the cost of this procedure is equal to the 
cost of $M_a$ (which is equal to that of $M_b$).  We conclude that
\begin{equation}
C(M^{(8)}_a) \le C(M_a).
\end{equation}
As we will see shortly, the cost of $M^{(8)}_a$ is in fact strictly smaller for $0<a<1$.

The measurement $M^{(8)}_a$ is a non-orthogonal measurement, but any non-orthogonal 
measurement can be performed by preparing an auxiliary system in a known
state and then performing a global orthogonal measurement on the combined system. 
We now show explicitly how to perform this particular measurement, in a way that
will allow us to determine the value of $C(M^{(8)}_a)$.  To do the measurement, 
Alice and Bob draw, from their store of entanglement, the entangled state
$|\phi^+_a\rangle = a|00\rangle + b|11\rangle$ of qubits C and D.  (As always, Alice holds C and Bob holds D.)  
Then each of them locally performs the Bell measurement $\{|\Phi^+\rangle, |\Phi^-\rangle,
|\Psi^+\rangle, |\Psi^-\rangle\}$ on his or her pair of qubits. 

The resulting 16-outcome orthogonal measurement on ABCD defines a 16-outcome POVM on just 
the two qubits A and B.  For each outcome $k$ of the global orthogonal measurement, we can find
the corresponding POVM element $\Pi_k$ of the AB measurement as follows:
\begin{equation}
\Pi_k = \hbox{tr}_{CD}\{ \pi_k [I_{AB} \otimes (|\phi^+_a\rangle\langle\phi^+_a|)_{CD}] \},
\end{equation}
where $\pi_k$ is the $k$th POVM element of the global measurement.  Less formally,
we can achieve the same result by taking the ``partial inner product'' between 
the initial state $|\phi^+_a\rangle$ of the system CD and the $k$th eigenstate of the global
measurement.  For example, the eigenstate $|\Phi^+\rangle|\Phi^+\rangle$ yields
the following partial inner product:
\begin{equation}
(\langle \phi^+_a|_{CD})(|\Phi^+\rangle_{AC}|\Phi^+\rangle_{BD}),
\end{equation}
which works out to be $(1/2)|\phi^+_a\rangle_{AB}$.  The corresponding POVM element
on the AB system is $(1/4)|\phi^+_a\rangle\langle\phi^+_a|$.  Continuing in this way,
one finds the following correspondence between the 16 outcomes of the global measurement and the POVM
elements of the AB measurement.
\begin{equation}
\begin{split}
&|\Phi^+\rangle|\Phi^+\rangle\hspace{2mm} \hbox{or}\hspace{2mm} |\Phi^-\rangle|\Phi^-\rangle\hspace{2mm}\rightarrow \hspace{2mm}\frac{1}{4}|\phi^+_a\rangle
\langle\phi^+_a|  \\
&|\Phi^+\rangle|\Phi^-\rangle\hspace{2mm} \hbox{or}\hspace{2mm} |\Phi^-\rangle|\Phi^+\rangle\hspace{2mm}\rightarrow \hspace{2mm}\frac{1}{4}|\phi^-_b\rangle
\langle\phi^-_b|  \\
&|\Phi^+\rangle|\Psi^+\rangle\hspace{2mm} \hbox{or}\hspace{2mm} |\Phi^-\rangle|\Psi^-\rangle\hspace{2mm}\rightarrow \hspace{2mm}\frac{1}{4}|\psi^+_a\rangle
\langle\psi^+_a|  \\
&|\Phi^+\rangle|\Psi^-\rangle\hspace{2mm} \hbox{or}\hspace{2mm} |\Phi^-\rangle|\Psi^+\rangle\hspace{2mm}\rightarrow \hspace{2mm}\frac{1}{4}|\psi^-_b\rangle
\langle\psi^-_b|  \\
&|\Psi^+\rangle|\Psi^+\rangle\hspace{2mm} \hbox{or}\hspace{2mm} |\Psi^-\rangle|\Psi^-\rangle\hspace{2mm}\rightarrow \hspace{2mm}\frac{1}{4}|\phi^+_b\rangle
\langle\phi^+_b|  \\
&|\Psi^+\rangle|\Psi^-\rangle\hspace{2mm} \hbox{or}\hspace{2mm} |\Psi^-\rangle|\Psi^+\rangle\hspace{2mm}\rightarrow \hspace{2mm}\frac{1}{4}|\phi^-_a\rangle
\langle\phi^-_a|  \\
&|\Psi^+\rangle|\Phi^+\rangle\hspace{2mm} \hbox{or}\hspace{2mm} |\Psi^-\rangle|\Phi^-\rangle\hspace{2mm}\rightarrow \hspace{2mm}\frac{1}{4}|\psi^+_b\rangle
\langle\psi^+_b|  \\
&|\Psi^+\rangle|\Phi^-\rangle\hspace{2mm} \hbox{or}\hspace{2mm} |\Psi^-\rangle|\Phi^+\rangle\hspace{2mm}\rightarrow \hspace{2mm}\frac{1}{4}|\psi^-_a\rangle
\langle\psi^-_a| 
\end{split}
\end{equation}
Thus, even though there are formally 16 outcomes of the AB measurement, they are equal in pairs, so that there are only eight distinct outcomes, and they are indeed the outcomes of 
the measurement $M^{(8)}_a$.  

The cost of this procedure is ${\mathcal E}(|\phi^+_a\rangle) = h(a^2)$.  This is the 
same as the average entanglement of the eight states representing the outcomes of $M^{(8)}_a$, 
which we know is a lower bound on the cost.  Thus the lower bound is achievable in this case,
and we can conclude that $C(M^{(8)}_a)$ is exactly equal to $h(a^2)$.  

We note that the POVM $M^{(8)}_a$ is invariant under all local Pauli operations.  This fact leads
us to ask whether, more generally, invariance under such operations guarantees that 
the entanglement cost of the measurement is exactly equal to the average entanglement
of the states associated with the POVM elements.  The next section shows that this is 
indeed the case for complete POVMs.  

\subsection{2.  An arbitrary complete POVM invariant under local Pauli operations}

We begin by considering a POVM on a bipartite system of dimension $d \times d$,
generated by applying generalized Pauli operators to a single pure state $|\phi_0\rangle$.
The POVM elements are of the form $(1/d^2)|\psi_{j_1k_1j_2k_2}\rangle\langle\psi_{j_1k_1j_2k_2}|$, where
\begin{equation}
|\psi_{j_1k_1j_2k_2}\rangle = (Z^{j_1}X^{k_1} \otimes Z^{j_2}X^{k_2}) |\phi_0\rangle
\label{genPOVM}
\end{equation}
and each index runs from $0$ to $d-1$.
Here the generalized Pauli operators $X$ and $Z$ are defined by
\begin{equation}
X|m\rangle = |m+1\rangle,  \hspace{1mm} Z|m\rangle = \omega^m |m\rangle, \hspace{1mm}
m = 0, \ldots, d-1,
\end{equation}
with $\omega = \exp(2\pi i/d)$ and with the addition understood 
to be mod $d$.  One can verify that the above construction generates a 
POVM for any choice of $|\phi_0\rangle$.  

In order to carry out this POVM, Alice and Bob use, as a resource, particles
C and D in the 
state $|\phi_0\rangle^*$, which has the same entanglement as $|\phi_0\rangle$.
(As before, the asterisk indicates complex conjugation in the standard basis.)
Alice performs on AC, and Bob on BD, the generalized Bell measurement whose
eigenstates are
\begin{equation}
|B_{jk}\rangle = \frac{1}{\sqrt{d}}Z^j\otimes X^k\sum_{r=0}^{d-1}|r,r\rangle.
\end{equation}

To see that this method does effect the desired POVM, we compute the partial inner
products as in the preceding subsection:
\begin{equation}
\begin{split}
&\left(\langle \phi_0|_{CD}^*\right) \left( |B_{j_1k_1}\rangle_{AC} \otimes |B_{j_2k_2}\rangle_{BD} \right)\\
&=\frac{1}{d}\sum_{r_1,r_2}\langle\phi_0|_{CD}^*Z_A^{j_1}X_C^{k_1}Z_B^{j_2}X_D^{k_2}|r_1,r_1\rangle_{AC}|r_2,r_2\rangle_{BD}\\
&= \frac{1}{d}\sum_{r_1,r_2}Z_A^{j_1}Z_B^{j_2}|r_1,r_2\rangle_{AB}\langle \phi_0|_{CD}^*X_C^{k_1}X_D^{k_2}|r_1,r_2\rangle_{CD}\\
&=\frac{1}{d}Z_A^{j_1}Z_B^{j_2}\sum_{r_1,r_2}|r_1,r_2\rangle_{AB}
\langle r_1,r_2|_{CD} X_C^{k_1}X_D^{k_2}|\phi_0\rangle_{CD}\\
&=\frac{1}{d}Z_A^{j_1}Z_B^{j_2}X_A^{k_1}X_B^{k_2}|\phi_0\rangle_{AB}\\
&=\frac{1}{d}(Z^{j_1}X^{k_1} \otimes Z^{j_2}X^{k_2}) |\phi_0\rangle_{AB}.
\end{split}
\end{equation}
Thus the combination of Bell measurements yields the POVM defined by Eq.~(\ref{genPOVM}).

We now extend this example to obtain the following result.

\medskip

\noindent {\bf Proposition 6.} Let $M$ be any complete POVM with a finite number of outcomes, acting on a 
pair of systems each having a $d$-dimensional state space, such that $M$ is
invariant under all local Pauli operations, that is, under the group generated by 
$X\otimes I$, $Z\otimes I$, $I\otimes X$, and $I\otimes Z$.  Then $C(M)$ is equal to 
the average entanglement of the states associated with the outcomes of $M$, as expressed
in Eq.~(\ref{ave}).

\medskip

\noindent {\em Proof.}  The most general such POVM is similar to the 
one we have just considered, except that instead of a single starting state $|\phi_0\rangle$,
there may be an ensemble of states $|\phi_s\rangle$ with weights $p_s$, $s=1,\ldots, m$,
such that $\sum p_s = 1$.  The POVM elements (of which there are a total of $md^4$) are 
$(p_s/d^2)|\psi_{j_1k_1j_2k_2;s}\rangle\langle\psi_{j_1k_1j_2k_2;s}|$, where
\begin{equation}
|\psi_{j_1k_1j_2k_2;s}\rangle = (Z^{j_1}X^{k_1} \otimes Z^{j_2}X^{k_2}) |\phi_s\rangle.
\end{equation}
(So $p_s/d^2$ plays the role of $\alpha_i$ in Eq.~(\ref{ave}).)
In order to perform this measurement, Alice and Bob first make a random choice
of the value of $s$, using the weights $p_s$.  They then use, as a resource, particles
C and D in the state $|\phi_s\rangle^*$, and perform Bell measurements as above.
The cost of this procedure is the average entanglement of the resource states, which is
\begin{equation}
\begin{split}
\hbox{cost} &= \sum_s p_s {\mathcal E}(|\psi_s\rangle) \\
&= \frac{1}{d^2}\sum_{j_1k_1j_2k_2s} \frac{p_s}{d^2} {\mathcal E}(|\phi_{j_1k_1j_2k_2;s}\rangle) \\
&=\langle {\mathcal E} \rangle .
\end{split}
\end{equation}
But we know that $\langle {\mathcal E} \rangle$ is a lower bound on $C(M)$.  Since the above
procedure achieves this bound, we have that 
$C(M) = \langle {\mathcal E} \rangle$. \hfill $\square$

\section{VII.  Discussion}

As we discussed in the Introduction, a general lower bound on the entanglement
cost of a complete measurement is the average entanglement of the pure states
associated with the measurement's outcomes.  Perhaps the most interesting
result of this paper is that, for almost all the orthogonal measurements we considered,
the actual cost is strictly greater than this lower bound.  The same is true
in the examples of ``nonlocality without entanglement'', in which the average
entanglement is zero but the cost is strictly positive.  However, whereas those earlier examples may have 
seemed special because of their intricate construction, the examples given
here are quite simple.  
The fact that the cost in these simple cases exceeds the average entanglement of the states
suggests that this feature may be a generic property of bipartite measurements.  
If this is true, then in this sense the nonseparability of a measurement is 
generically a distinct property from the the nonseparability of the eigenstates.  
(In this connection it is interesting that for certain questions of distinguishability
of generic bipartite states, the presence or absence of entanglement seems to be completely
irrelevant \cite{WalgateScott}.)

We have also found a class of measurements for which the entanglement cost is {\em equal to}
the average entanglement of the corresponding states.  These measurements have a high degree of
symmetry in that they are invariant under all local generalized Pauli operations.  

What is it that causes some measurements to be ``more nonseparable'' than the states
associated with their outcomes?  Evidently the answer must have to do with the {\em relationships} among the
states.  In the original ``nonlocality without entanglement'' measurement, the crucial
role of these relationships is clear: in order to separate any 
eigenstate $|v\rangle$ from any other eigenstate $|w\rangle$ by a local measurement, the observer must disturb some of the other states
in such a way as to render them indistinguishable.  One would like to have a similar understanding
of the ``interactions'' among states when the eigenstates are entangled. 
Some recent papers
have quantified relational properties 
of ensembles of bipartite states \cite{Horodecki2, Horodecki3}. 
Perhaps one of these approaches, or a different approach yet to be developed,
will capture the aspect of these relationships that determines the cost of the
measurement.  

\begin{acknowledgments}
We thank Alexei Kitaev, Debbie Leung, David Poulin, John Preskill, 
Andrew Scott and Jon Walgate for valuable discussions and comments on the subject.
S.\,B.~is supported by Canada's Natural Sciences and Engineering Research Council
({\sc Nserc}).
G.\,B.~is supported by Canada's Natural
Sciences and Engineering Research Council ({\sc Nserc}),
the Canada Research Chair program,
the Canadian Institute for Advanced Research ({\sc Cifar}),
the Quantum\emph{Works} Network and the Institut transdisciplinaire d'informatique quantique (\textsc{Intriq}).

\end{acknowledgments}

\appendix*
\section{Appendix: The one-round cost in the limit of small entanglement}

\noindent {\em Lower bound}

Our lower bound on the one-round cost is given by Eqs.~(\ref{lowerbound1}) and (\ref{lowerbound2}),
which we rewrite here in an equivalent form:
\begin{equation}
\hbox{cost} \ge \frac{(ac+bd)^2 - c^2}{d^2},   \label{firstapp}
\end{equation}
where $d$ is determined by the equation
\begin{equation}
h\left[\frac{(ac+bd)^2-c^2}{(ac+bd)^2}\right]=\frac{(ac+bd)^2 - c^2}{d^2}.  \label{appequation}
\end{equation}
For a small value of the parameter $b$, we would like to obtain an approximation to the
value of $d$ that solves Eq.~(\ref{appequation}).  As discussed in Section III, we are
looking for a solution in the range $b \le d \le 1/\sqrt{2}$, and the forms of the functions
in Eq.~(\ref{appequation}) guarantee that there will be a unique solution in this range.
One can show that within this range, the right-hand
side of Eq.~(\ref{appequation}) satisfies the inequalities
\begin{equation}
\frac{b}{d} \le \frac{(ac+bd)^2 - c^2}{d^2} \le \frac{2b}{d}. \label{firstineq}
\end{equation}
Applying these same inequalities to the argument of the function $h$ on the left hand
side of Eq.~(\ref{appequation}), we have
\begin{equation}
\frac{bd}{(ac+bd)^2} \le \frac{(ac+bd)^2-c^2}{(ac+bd)^2} \le \frac{2bd}{(ac+bd)^2}.  \label{app2}
\end{equation}
For sufficiently small $b$, the function $h$ evaluated at the values
appearing in 
Eq.~(\ref{app2}) is an increasing function, so we can write
\begin{equation}
h\left[\frac{bd}{(ac+bd)^2}\right] \le h\left[\frac{(ac+bd)^2-c^2}{(ac+bd)^2}\right]
\le h\left[\frac{2bd}{(ac+bd)^2}\right].    \label{explicith}
\end{equation}
We can bound the entropies to obtain
\begin{equation}
-bd\log b \le h\left[\frac{(ac+bd)^2-c^2}{(ac+bd)^2}\right] \le -16 bd\log b.  \label{entest}
\end{equation}
 Combining Eqs.~(\ref{appequation}), (\ref{firstineq}), and (\ref{entest}), we get
 \begin{equation}
 -\frac{1}{2} \log b \le \frac{1}{d^2} \le -16 \log b.
 \end{equation}
 Thus $d$ goes to zero as $b$ goes to zero, but it does so much more slowly.  
 
 We now use this observation to approximate each side of Eq.~(\ref{appequation}).
 First, in the entropy function $h(x) = -x\log x - (1-x) \log (1-x)$, for very small $x$
 we can ignore the second term, so that Eq.~(\ref{appequation}) can be
 simplified to 
 \begin{equation}
 -\log\left[\frac{(ac+bd)^2-c^2}{(ac+bd)^2}\right] \approx \frac{(ac+bd)^2}{d^2}.
 \end{equation}
 Now, with $b$ very small and $d$ of order $1/\sqrt{\log(1/b)}$, we can approximate
 $(ac+bd)^2 - c^2$ as
 \begin{equation}
 (ac+bd)^2 - c^2  \approx 2bd.
 \end{equation}
 So the equation becomes $ -\log(2bd) \approx \frac{1}{d^2}$, but since
 $-\log 2d$ becomes negligible compared to $-\log b$, we can just as well write
 \begin{equation}
 -\log b \approx \frac{1}{d^2}.
 \end{equation}
 Finally, the lower bound given by Eq.~(\ref{firstapp}) becomes
 \begin{equation}
 \hbox{lower bound}\; \approx \frac{2bd}{d^2} \approx 2b\sqrt{\log(1/b)}.
 \end{equation}

All of our approximations have been such that the ratio between the approximating
function and the exact function approaches unity as $b$ approaches zero.  So 
the same is true of the approximate expression $2b\sqrt{\log(1/b)}$ relative
to the exact lower bound.  

\bigskip

\noindent {\em Upper bound}

Our upper bound for the single-round cost (Eq.~(\ref{oneround})) is the minimum over $q$ in the range
$0 \le q \le 1/2$ of the function
\begin{equation}
f(q) = h(q) + g(q),
\end{equation}
where 
\begin{equation}
g(q) = 1 - \frac{1}{(a^2/(1-q)) + (b^2/q)}.
\end{equation}
(Here $q$ is playing the role of $y^2$ in Eq.~(\ref{oneround}).)  The function $g(q)$ decreases
monotonically from the value $1$ at $q=0$ to its minimum value $2ab/(1+2ab)$ at
$q=b/(a+b)$.  Thus the minimum value of $g(q)$ approaches zero for small $b$ 
and is attained arbitrarily close to $q=0$.  Therefore for sufficiently small $b$, the 
function $g(q)$, as it falls to its minimum value, falls farther 
than $h(q)$ rises, and the minimum value of $f(q)$ 
is less than 1.  This minimum value is attained at some value of $q$---call it
$q_0$---which is less than $b/(a+b)$.  (Beyond that point both $h(q)$ and
$g(q)$ are increasing for $q<1/2$.)  More simply, $q_0 < b$.  So we can 
limit our attention to values of $q$ less than $b$.

With this limitation, for small $b$ we can approximate the function $f(q)$ as
\begin{equation}
f(q) \approx -q\log q + \frac{q^2 + b^2}{q + b^2}.
\end{equation}
Setting the derivative of this function equal to zero, we find that $q_0$
can be made arbitrarily close (in the sense that the fractional error can be made 
arbitrarily small) to a solution of
\begin{equation}
\frac{b^2}{(q+b^2)^2} = -\log q.
\end{equation}
For small $b$ there are two solutions to this equation with $q<b$.  The smaller one,
with $q$ of order $\exp(-1/b^2)$, corresponds to a local {\em maximum}
of $f(q)$, reflecting the fact that the slope of $h(q)$ approaches positive infinity as $q$
approaches zero,
whereas the competing negative slope of $g(q)$ is finite at $q=0$.  The other solution,
with $q$ approximately equal to $b/\sqrt{\log(1/b)}$, is therefore the one we want.  
At this value we have 
$f(q) \approx 2b\sqrt{\log(1/b)}$.  Again, the approximation is such that the ratio of 
the exact upper bound to this approximate value approaches unity
as $b$ approaches zero.

\end{document}